\pdfoutput=1
\documentclass[sigconf]{acmart}

\usepackage{algorithm}
\usepackage{algpseudocode}
\usepackage{multirow}
\usepackage{amsthm}
\usepackage{amsmath}
\usepackage{subfig}
\newtheorem{definition}{Definition}
\newtheorem{problem}{Problem}

\begin{document}

\title{Detection of Crowdsourcing Cryptocurrency Laundering via Multi-Task Collaboration}



\author{Guang Li}
\affiliation{%
  \institution{Sun yat-sen University}
  \city{Guangzhou}
  \country{China}}
\email{liguang7@mail2.sysu.edu.cn}

\author{Litong Sun}
\affiliation{%
  \institution{Sun yat-sen University}
  \city{Guangzhou}
  \country{China}}
\email{Sunlt5@mail2.sysu.edu.cn}

\author{Jieying Zhou}
\affiliation{%
  \institution{Sun yat-sen University}
  \city{Guangzhou}
  \country{China}}
\email{isszjy@mail.sysu.edu.cn}

\author{Weigang Wu}
\affiliation{%
  \institution{Sun yat-sen University}
  \city{Guangzhou}
  \country{China}}
\email{wuweig@mail.sysu.edu.cn}







\begin{abstract}
  USDT, a stablecoin pegged to dollar, has become a preferred choice for money laundering due to its stability, anonymity, and ease of use. Notably, a new form of money laundering on stablecoins -- we refer to as crowdsourcing laundering -- disperses funds through recruiting a large number of ordinary individuals, and has rapidly emerged as a significant threat. However, due to the refined division of labor, crowdsourcing laundering transactions exhibit diverse patterns and a polycentric structure, posing significant challenges for detection. In this paper, we introduce transaction group as auxiliary information, and propose the Multi-Task Collaborative Crowdsourcing Laundering Detection (MCCLD) framework. MCCLD employs an end-to-end graph neural network to realize collaboration between laundering transaction detection and transaction group detection tasks, enhancing detection performance on diverse patterns within crowdsourcing laundering group. These two tasks are jointly optimized through a shared classifier, with a shared feature encoder that fuses multi-level feature embeddings to provide rich transaction semantics and potential group information. Extensive experiments on both crowdsourcing and general laundering demonstrate MCCLD’s effectiveness and generalization. To the best of our knowledge, this is the first work on crowdsourcing laundering detection.

  
\end{abstract}


\maketitle

\section{Introduction}

USDT, or Tether, is a popular stablecoin – a cryptocurrency pegged to and backed by fiat currencies like the U.S. dollar – which has the most liquid markets with an estimated US \$15 billion daily trading volume at the time of writing\footnote{https://coinmarketcap.com/currencies/tether/}. USDT on the TRON blockchain has become a preferred choice for regional cyberfraud operations and money launderers alike due to its stability, anonymity, low fees, and ease of use\cite{UNODC2024}. According to the 2025 Crypto Crime Report by Bitrace, a leading blockchain security firm, illicit transactions constituted 5.14\% of all stablecoin transactions, totaling approximately \$649 billion\cite{Bitrace2025}. Notably, stablecoin inflows to money laundering addresses accounted for \$86.3 billion, representing 13\% of total illicit transactions. Money laundering, the process of legitimizing criminal funds, is essential for illicit activities. Therefore, effective cryptocurrency laundering detection is crucial to combating crypto crimes.

\begin{figure}[!t]
\centering
\includegraphics[width=0.35\textwidth]{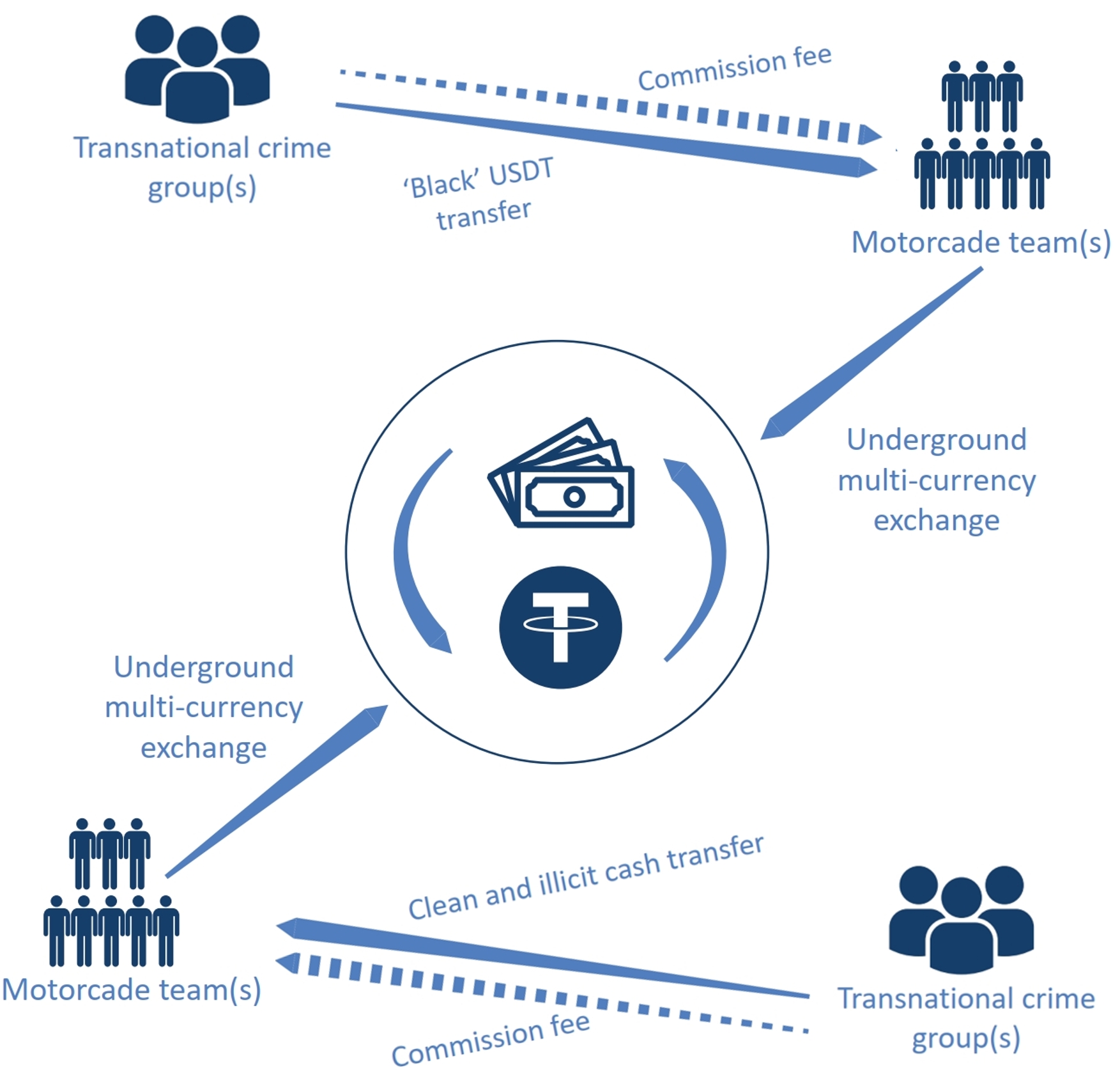}
\caption{Simplified USDT to fiat "motorcade" model\cite{UNODC2024}}
\label{fig:runningpoints}
\end{figure}  

Concerningly, a new form of money laundering called "running points" has emerged on stablecoins. This scheme recruits individuals (called 'Motorcade') into money laundering networks and facilitates pass-through activities via social network APPs (like Telegram), posing a significant threat to financial security\cite{UNODC2024}. Fig. \ref{fig:runningpoints} illustrates this laundering model. In practice, motorcade teams typically adopt multi-level pyramid structures with tiered bonuses and responsibilities, including multi-level agents and bottem-level laundering labor. Similar activities also include "black USDT sales", which launder money by selling discounted dirty USDT to recruited individuals, subsequently transferring an equivalent amount of clean funds to designated accounts. 

\begin{figure}[!t]
\centering
\subfloat[Hacker laundering]{\includegraphics[width=0.2\textwidth]{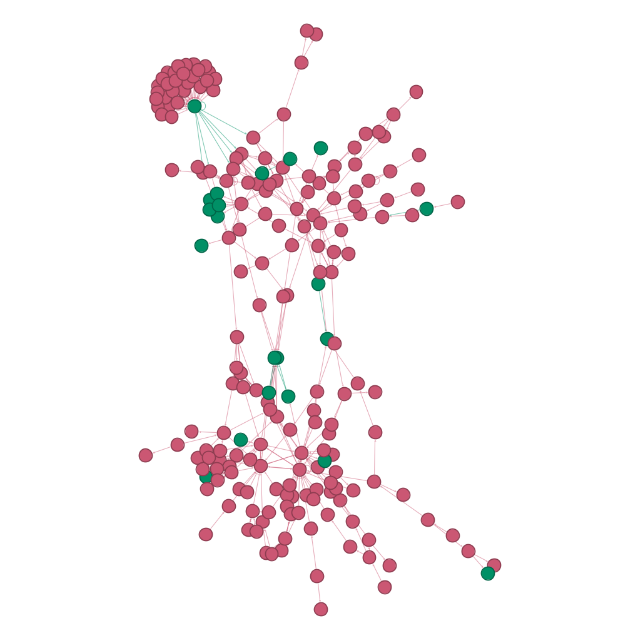}
\label{fig:hacker_laundering}}
\hfil
\subfloat[Crowdsourcing laundering]{\includegraphics[width=0.2\textwidth]{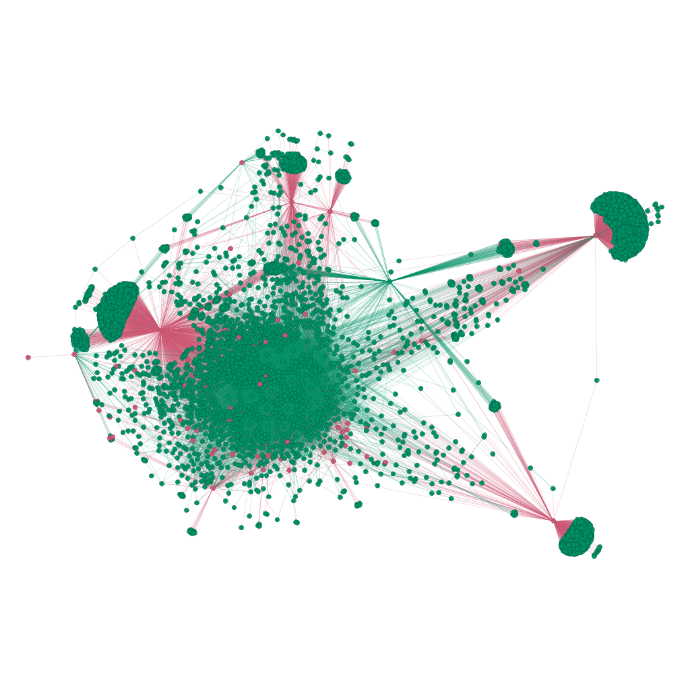}
\label{fig:crowd_laundering}}
\caption{Examples of transaction subgraphs for two types of money laundering}
\label{fig:difference}
\end{figure}

\begin{figure}[!t]
\centering
\includegraphics[width=0.4\textwidth]{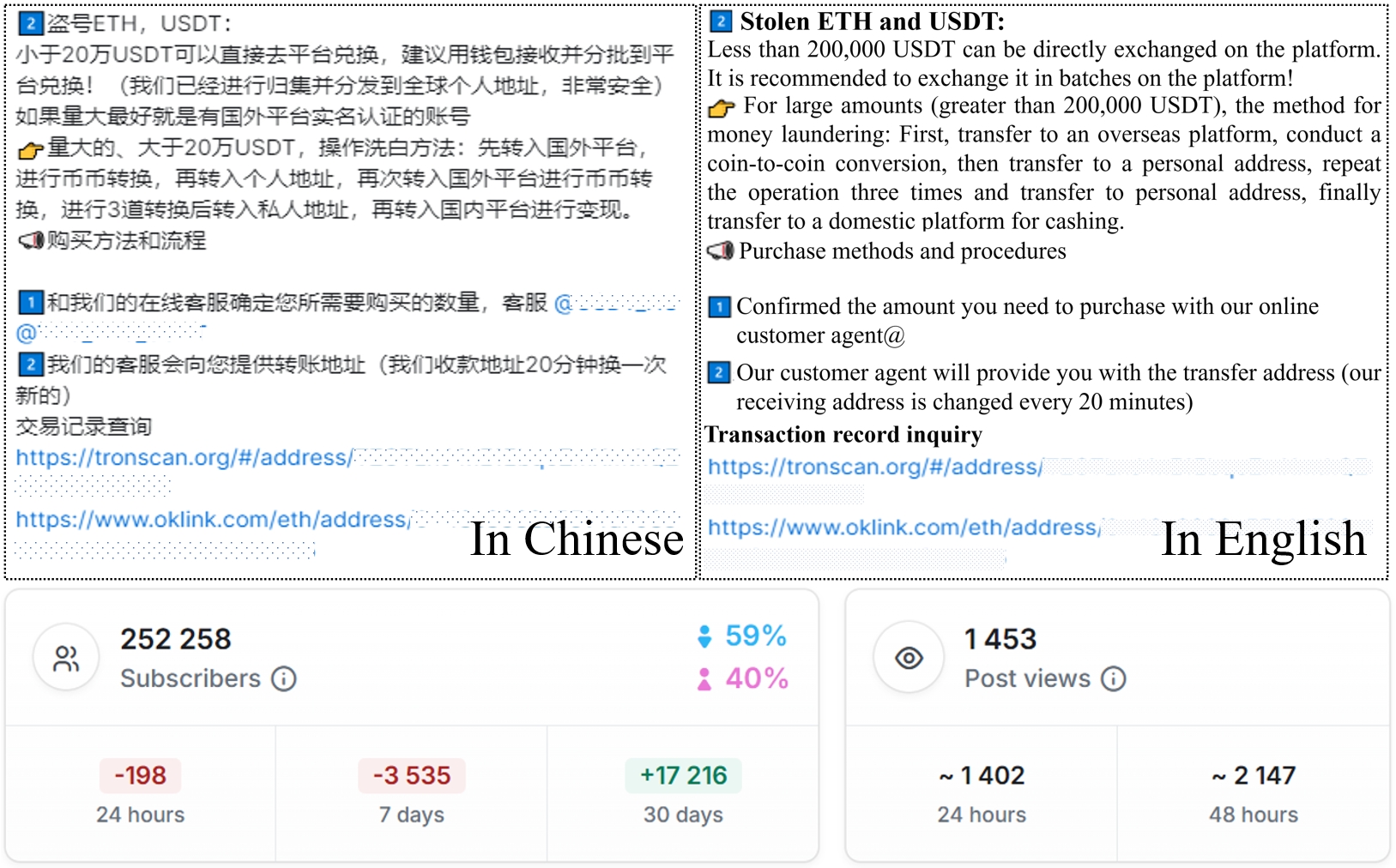}
\caption{A chat announcement in Telegram}
\label{fig:heitransaction}
\end{figure} 


Such crowdsourcing laundering brings new challenges different from those of general cryptocurrency laundering (e.g., hacker laundering). First, general laundering process is typically performed by criminals themselves, where controlled accounts exhibit similar transaction patterns. However, crowdsourcing laundering involves numerous agents and labor -- usually ordinary individuals who pass exchange KYC checks, and exhibit significantly different transaction patterns. This raises generalization challenges for detection. Second, general laundering transactions are usually centralized within a gang. In contrast, in crowdsourcing laundering, the refined division of labor and sharing labor groups across platforms create highly interactive multi-gang activities, raising the challenge of detecting complete laundering chains. As depicted in Fig. \ref{fig:difference}, the hacker laundering subgraph (Fig. \ref{fig:hacker_laundering}; Harmony hack incident\footnote{https://x.com/harmonyprotocol/status/1540110924400324608}) is dominated by internal laundering accounts, exhibiting a layered structure with limited node scale. Conversely, the crowdsourcing laundering subgraph (Fig. \ref{fig:crowd_laundering}) mixes multiple gangs in a polycentric structure with a large number of uncategorized accounts. At a high level, traditional cryptocurrency laundering can be viewed as a simplified form of crowdsourcing laundering.

Existing methods focus on either laundering individual or subgraph detection. Individual-based methods assume laundering entities are abnormal to normal entities, employing supervised classifiers like XGBoost, RandomForest, Graph convolutional networks to build detection models\cite{Weber2019, Elmougy2023, Ranshous2017, Hu2019, Dylan2021}. However, due to the refined division of labor, transaction pattern of different labor are significantly different in crowdsourcing laundering, limiting traditional classifiers' generalization; Subgraph-based methods consider money laundering as a group behavior with collaborative of multiple accounts, and detect such activities from the perspective of subgraph by technologies like pattern matching\cite{Lee2020,Starnini2021}, densest subgraph discovery\cite{Li2020,Lin2024}, and graph partition\cite{Cheng2023,Chai2023,LiG2025}. However, crowdsourcing laundering is a more complex collaboration between multiple gangs, these methods are hard to capture the full picture of the polycentric structure generated by multi-group collaboration.

To address these challenges, we introduce transaction groups as auxiliary information and propose the Multi-Task Collaborative Crowdsourcing Laundering Detection (MCCLD) framework. MCCLD employs an end-to-end graph neural network with shared feature extraction and classifiers to collaborate on transaction group detection and laundering transaction detection tasks. For feature sharing, we extract and fuse multi-level embeddings (account, transaction and group) from transaction graphs, providing rich transaction semantics and potential group information. For classifier sharing, we first transform the transaction group detection task, which is typically a clustering problem, into a binary edge classification model, where each edge is classified as either intra-group (1) or inter-group (0). This transformation preserves transaction group connectivity while maintaining the classification framework for laundering detection. Then, a unified classifier is constructed to jointly optimize the two tasks, targeting at learning discriminative laundering features while reducing differences in transaction embeddings within the same group. Finally, we achieve detection on crowdsourcing laundering transactions by this effective collaborative mechanism.

The main contributions of this paper are summarized as follows: 

\begin{enumerate}
    \item \textbf{Methodology.} We propose a multi-task collaborative crowdsourcing laundering detection (MCCLD) framework for detecting crowdsourcing laundering transactions with internal pattern variations and polycentric structures. To the best of our knowledge, this is the first work on crowdsourcing laundering detection. 

    \item \textbf{Dataset.} We manually collect crowdsourcing laundering cases from Telegram chats (e.g., Fig. \ref{fig:heitransaction}) and validate relevant addresses \footnote{https://drive.google.com/drive/folders/1fxpc17tNKSebCuE3nJBqcmwkinD2jTAb} via public annotations on blockchain explorers. Moreover, we discover that the mechanism of resource pledge and delegation in TRON is usually used to reduce transaction fees in crowdsourcing laundering, providing ideal supervised group information for MCCLD to bridge different groups. 

    \item \textbf{Experimentation.} We evaluate MCCLD on both crowdsourcing and general laundering detection. Results show MCCLD not only effectively detects crowdsourcing laundering transactions but also achieves strong generalization performance on general laundering detection. Specifically, MCCLD outperforms existing methods by 53.4\% on average in crowdsourcing laundering detection and 36.9\% in general laundering detection. Moreover, benefiting from group information and GNN's semi-supervised nature, MCCLD maintains 70\% detection accuracy with only 10\% labeled data, demonstrating practical advantages.
\end{enumerate}

\section{Related Works}
This section reviews recent money laundering detection methods on cryptocurrencies. These methods are mainly divided into two categories based on the target to detect: individual-based detection and subgraph-based detection.

\subsection{Individual-based Laundering Detection}

The core objective of individual-based detection is to precisely distinguish laundering entities from normal transaction entities. Initially, detection primarily relied on expert knowledge\cite{rajput2014} and regulatory guidelines outlining money laundering characteristics\cite{Khanuja2014} to build Anti-Money Laundering (AML) rule repositories. These rule-based methods retain significant industry value due to high interpretability, regulatory compliance, and deployment ease. Sepecifically, The Financial Action Task Force (FATF) released travel rules \cite{FATFTRS2025} in 2025 to enable countries to better establish cryptocurrency laundering supervision in practice. Bitquery, a prominent crypto analytics firm, developed a series of key laundering indicators based on FATF red flags\footnote{https://www.bitquery.io/blog/fatf-indicators-bitquery-detect-money-laundering}. However, these methods are often limited by the complexity of laundering behaviors and the dynamic nature of laundering tactics, leading to high false positive rates and low detection accuracy.

Driven by advancements in machine learning and graph neural networks (GNNs), AML technologies have transitioned to data-driven approaches. These methods extract features from transaction data, using labeled data for supervised training to build classification models, ultimately achieving the automated recognition of laundering entities. Sepecifically, Dylan et al. \cite{Dylan2021} developed an ensemble method that detect laundering via data stream analysis. Lorenz et al. \cite{Lorenz2021} introduced active learning to address the issue of label scarcity in cryptocurrency laundering detection. Weber et al. \cite{Weber2019} utilized GCN and Evolve-GCN to detect laundering transactions in the Bitcoin network. Alarab et al. \cite{Cardoso2022} proposed a relational-GCN to enhance node interrelations within node embeddings and improve the laundering detection accuracy. The performance of these methods is mainly attributed to two aspects. 1) Classifiers, various machine learning technologies have been employed to improve classification accuracy, including ensemble learning \cite{Hu2019,Dylan2021}, active learning \cite{Lorenz2021} and GNNs \cite{Weber2019,Cardoso2022}. 2) Features, various features have been developed to train classifiers, including transaction attributes, graph statistical features, and graph embedding features. Note that almost all methods show that graph embedding features are more effective than other features \cite{Hu2019,Weber2019,Cardoso2022,Elmougy2023,oliveira2021}. 

While effective for laundering individual detection, these methods target specific laundering individuals. They struggle with crowdsourcing laundering where different labor exhibit significantly varied transaction patterns, leading to poor generalization. In this paper, we address this by introducing transaction group detection as an auxiliary task to reduce intra-group embedding differences, enhancing detection generalization.


\subsection{Subgraph-based Laundering Detection}
Subgraph-based methods try to detect laundering by recognizing subgraph patterns like layered structures \cite{Starnini2021,Lee2020} and dense transfers \cite{Li2020,Sun2022}, and techniques including densest subgraph discovery, and graph partition are employed. Densest subgraph discovery-based methods assume that laundering activities exhibit distinctively dense features in terms of certain metrics. Specically, Chen et al. \cite{Chen2022} proposed AntiBenford, combining Benford's Law with dense subgraph discovery to extract anomalous subgraphs via $\chi^{2}$-based node metrics. Lin et al. \cite{Lin2024} proposed DenseFlow, which designs suspiciousness indicators across topological, temporal, and other dimensions. Based on these indicators, it traces suspicious Ethereum laundering fund flows by using dense subgraph discovery and maximum flow algorithms. However, the effect of these methods is similar to that of heuristic algorithms, and the subgraphs they found are usually local solutions within networks.

Instead of directly mining laundering subgraphs, graph partition-based methods usually follow a two-phase paradigm, including subgraph partition and laundering detection. Specically, Li et al. \cite{LiG2025} proposed GMPA, which adopts an overlapping clustering to partition transaction network into account groups based on transaction personas, enabling laundering groups are isolated from normal transaction relationships. Based partitioned account groups, it detects laundering groups via cycle-based features (cycle basis number and overlapping ratio). However, the performance of graph partition-based method heavily depends on the quality of first phase, either the precision of subgraph partition or laundering detection.  

In summary, subgraph-based methods can effectively detect laundering subnetwork of inter-individual collaboration but ignore inter-group collaboration, and they are unable to capture complete crowdsourcing laundering activities in polycentric structure. To address this challenge, we introduce supervised transaction group information (e.g., resource pledges and delegations), to bridge between laundering groups and achieve detection on such polycentric laundering structure. 


\section{The Proposed Method}
\subsection{Multi-Task Collaboration Framework}

\begin{figure*}[!t]
\centering
\includegraphics[width=0.9\textwidth]{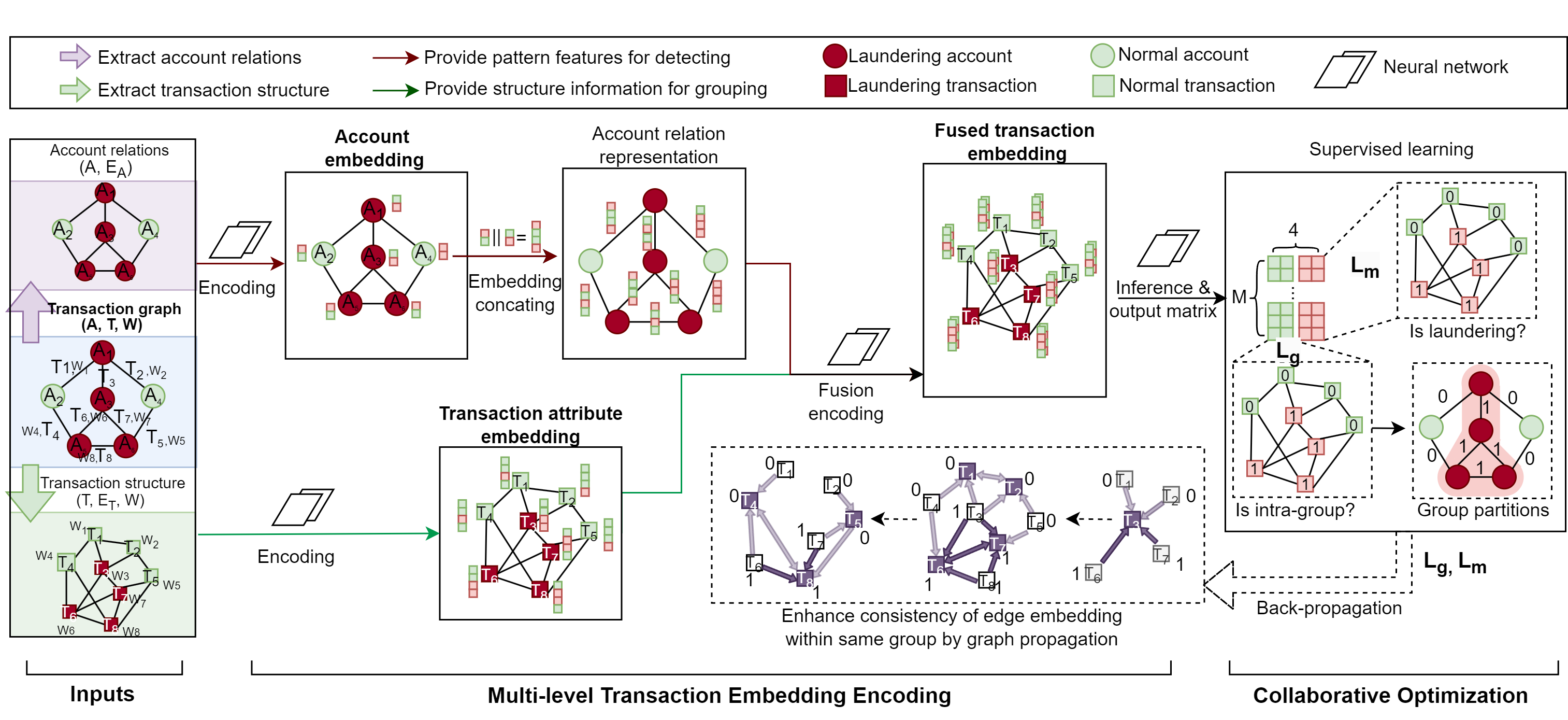}
\caption{Framework of the MCCD Algorithm}
\label{fig:crowdsourcingmlstructure}
\end{figure*}

Our proposed detection framework involves two tasks: laundering transaction detection and transaction group detection. The latter is introduced as an auxiliary mechanism to enhance the detection performance for crowdsourcing laundering transactions that exhibit diverse patterns and polycentric structure. 

However, there is heterogeneity between these two tasks: laundering transaction detection is typically modeled as a classification task, while transaction group detection is treated like a graph clustering or community detection problem. This heterogeneity leads to two bottlenecks in task collaborative. First, the number of categories in laundering transaction detection is fixed (typically binary classification), whereas the number of groups in transaction group detection varies with transaction networks. Second, transaction groups are typically connected components within transaction network, but it is challenging to ensure connectivity constraints for detection results in neural networks.


To address this heterogeneity, we align both tasks as edge-level binary classification. Specifically, we transform transaction group detection into a binary classification problem where each edge is classified as either intra-group (1) or inter-group (0). This transformation preserves the connectivity of transaction groups within the classification framework of money laundering detection. Thus, we indicate the partitioning of transaction groups by using Transaction Group Vector:

\begin{definition}[Transaction Group Vector]
  Given a transaction network \(\mathcal{G} = (\mathcal{A}, \mathcal{T}, \mathcal{W})\), let $\mathbf{C} = \{C_1, C_2, \ldots, C_k\}$ denote the set of ground-truth account groups, where $k$ is the total number of groups. Each account group \(C_i\) corresponds to a weakly connected component in \(\mathcal{G}\). The transaction group indicator vector $\mathbf{I} \in \{0,1\}^m$ is defined over the transaction set $\mathcal{T}$ (with $|\mathcal{T}| = m$). For each transaction $\mathcal{T}_i = (u,v)$:
    \begin{equation}
        \mathbf{I}_i = \left\{
        \begin{aligned}
            1 & \quad \text{if } \mathcal{T}_i \in C_j, j \in \{1,2,\ldots,k\} \\
            0 & \quad \text{otherwise}
        \end{aligned}
        \right. .
    \end{equation}
\end{definition}

Note that if $\mathcal{G}$ is fully partitioned into disconnected components, the Transaction Group Vector $\mathbf{I}$ would degenerate to an all-ones vector, losing discriminative power. Therefore, the transaction graph is constrained to only perform partial grouping in this paper. Further, the problem of transaction group detection is defined as:

\begin{problem}[Transaction Group Detection]\label{pro:groupdetection}
    Given a transaction network $\mathcal{G}$, and ground-truth transaction group vector $\mathbf{I}$:

    \begin{itemize}
        \item Find: A graph embedding extractor $\mathcal{F}_g$ and classifier $\mathcal{C}_g$
        \item Such that: For transactions $\mathcal{T}$, $\mathcal{C}_g(\mathcal{F}_g(\mathcal{G}, \mathcal{T})) \approx \mathbf{I}$ 
    \end{itemize}
    
\end{problem}

This binary classification modeling based on the transaction group vector inherently preserves connectivity: edges classified as 1 within a connected subgraph belong to the same group. Building on this, we define:

\begin{problem}[Laundering Transaction Detection]\label{pro:launderingdetection}
    Given a transaction network $\mathcal{G}$, and the laundering transaction labels $\mathbf{L}$:

    \begin{itemize}
        \item Find: A graph embedding extractor $\mathcal{F}_d$ and classifier $\mathcal{C}_l$
        
        \item Such that: For transactions $\mathcal{T}$, $\mathcal{C}_l(\mathcal{F}_d(\mathcal{G}, \mathcal{T})) \approx \mathbf{L}$
    \end{itemize}
    
\end{problem}

After aligning the problem modeling of the two tasks, we propose a framework to jointly optimize them, as illustrated in Fig. \ref{fig:crowdsourcingmlstructure}. Specically, in the MCCLD framework, the two tasks are collaboratively optimized in an end-to-end graph neural network, and aligned via a shared transaction graph embedding encoder (i.e., $\mathcal{F}_g = \mathcal{F}_d$) and a shared classifier (i.e., $\mathcal{C}_l = \mathcal{C}_g$). This shared embedding and decision-making mechanism enables the network to learn rich transaction embeddings that capture both of laundering transaction patterns and potential group information.

As illustrated in Fig. \ref{fig:crowdsourcingmlstructure}, the MCCLD framework extracts and fuses multi-level embedding features from the transaction graph, providing information-rich feature inputs for the classifier. Based on the shared classifier architecture, the network outputs a transaction detection matrix $\mathbf{M} \in \mathbb{R}^{m \times 4}$, where $m$ denotes the number of transactions and the output dimension per transaction is 4. The first two dimensions represent the one-hot encoding for money laundering transaction detection, while the last two dimensions represent the one-hot encoding for transaction group detection. Finally, the model is jointly optimized via supervised learning, achieving efficient collaboration between money laundering transaction detection and transaction group detection within a unified neural network.

\subsection{Multi-Level Transaction Encoding}
In this part, we focus on extracting semantically rich transaction embeddings $\mathbf{T}$ from the transaction network \(\mathcal{G}\). These embeddings serve as shared inputs for the unified classifier, simultaneously supporting both laundering transaction detection and transaction group detection. To satisfy the distinct requirements of these tasks, the embeddings must achieve two objectives: 1) characterizing transaction behavioral features precisely at attribute and relationship levels; 2) integrating intrinsic group representations to support grouping analysis. While GNNs provide an established pathway for encoding transaction attributes and structural features into \(\mathbf{T}\) via message passing, effectively capturing implicit group information remains a challenge in embedding encoding.


The groups, representing implicit structural information in cryptocurrency networks, are inferred from inter-account behaviors and denoted as account sets. Furthermore, transaction group refers to the set of relationships within account group. To this end, we obtain potential transaction group information through extracting graph embeddings from two levels, i.e., account level and transaction level. The account-level embeddings integrate account attributes and relationships, and the transaction-level embeddings encode transaction attributes and structure. As depicted in Fig. \ref{fig:crowdsourcingmlstructure}, we extract account and transaction relations from the transaction graph, encoding them into account embeddings and transaction embeddings respectively. This separate encoding enables direct representation and optimization of transactions, rather than indirect derivation from account embeddings (e.g., account relation representations in Fig. \ref{fig:crowdsourcingmlstructure}). Finally, we fuse these two levels of embeddings to form unified transaction embeddings.

Fig. \ref{fig:gnblock} illustrates the embedding fusion mechanism within the GNN block, where \(\phi\) denotes the update function, \(\rho\) denotes the aggregation function, and \(\theta\) represents the network parameters. The encoding process operates as follows:

\begin{figure}[!t]
\centering
\includegraphics[width=0.35\textwidth]{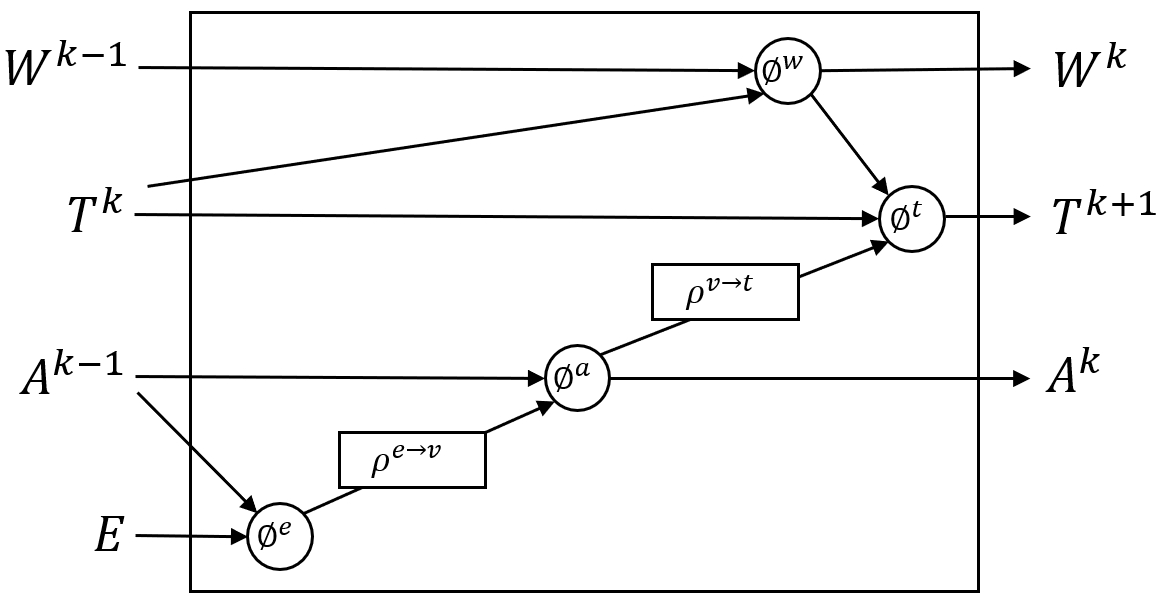}
\caption{Internal structure of GNN block}
\label{fig:gnblock}
\end{figure}

\begin{enumerate}
    \item \textbf{Account Embedding.} The \(k\)-th layer account embedding \(\mathbf{A}^{k}_{v_i}\) for account \(v_i\) is computed as:
    \begin{equation} \label{eq:accountembedding}
      \mathbf{A}^{k}_{v_i} = \phi^{a}_{\theta}\left(\mathbf{A}^{k-1}_{v_i}, M_{v_i}^{k}\right),
    \end{equation}
    where \(M_{v_i}^{k}\) is the aggregated neighbor message:  
    \begin{equation}
        \begin{aligned}
          M_{v_i}^{k} &= \rho^{e \rightarrow v}\left(\mathbf{A}^{k-1}, E_{v_i}\right) \\
          \rho^{e \rightarrow v}\left(\mathbf{A}^{k-1}, E_{v_i}\right) &= \frac{1}{|\mathcal{N}(v_i)|}\sum_{v_j \in \mathcal{N}(v_i)} \phi^{e}_{\theta}\left(\mathbf{A}^{k-1}_{v_j}\right).
        \end{aligned}
    \end{equation}
    \item \textbf{Transaction Attribute Embedding.} Transaction attributes are encoded via:  
    \begin{equation} \label{eq:weightembedding}
      \mathbf{W}^{k} = \phi^{w}_{\theta}\left(\mathbf{W}^{k-1}, \mathbf{T}^{k}\right).
    \end{equation}
    \item \textbf{Fused Transaction Embedding.} The fused transaction embedding for transaction $e_i=(v_{r_i}, v_{s_i})$ is obtained by:
    \begin{equation} \label{eq:transactionembedding}
      \mathbf{T}^{k+1}_{i} = \phi^{t}_{\theta}\left(\mathbf{T}^{k}_{i}, \left[\mathbf{W}^{k}_{e_i} \parallel M^{k}_{e_i}\right]\right), 
    \end{equation}
    where \(M^{k}_{e_i}\) is the aggregated transaction embedding result, and \(\left[\mathbf{W}^{k}_{e_i} \parallel M^{k}_{e_i}\right]\) denotes column-wise concatenation of transaction attribute and structural embeddings. \(M^{k}_{e_i}\) is computed as:  
    \begin{equation}
      M^{k}_{e_i} = \rho^{v \rightarrow t}\left(\mathbf{A}^{k}, {e_i}\right) = \left[\mathbf{A}^{k}_{v_{r_i}} \parallel \mathbf{A}^{k}_{v_{s_i}}\right].
    \end{equation}
\end{enumerate}

These encoding processes can be corresponded in Fig. \ref{fig:crowdsourcingmlstructure} as well. In practice, the update functions \(\phi^{e}\) and \(\phi^{a}\) are implemented using Graph Isomorphism Network (GIN) architecture, while \(\phi^{w}\) and \(\phi^{t}\) employ Multilayer Perceptron (MLP) architectures.

\subsection{Multi-Task Collaborative Optimization}
\begin{algorithm}[!t]
\caption{Multi-Task Collaborative Optimization}
\begin{algorithmic}[1]
\State \textbf{Input:} Transaction graph \(\mathcal{G}\), laundering transaction labels \(\mathbf{L}\), transaction group labels \(\mathbf{I}\)
\Statex
\For {each training epoch}
    \State Sample a batch of \(n\) subgraphs\(\{\mathcal{G}'_1, \dots, \mathcal{G}'_n\} \subset \mathcal{G}\)
    \For {each subgraph \(\mathcal{G}'_i\)}
        \State Compute fused embeddings \(\mathbf{T}\) via Eq. \ref{eq:accountembedding}, \ref{eq:weightembedding}, \ref{eq:transactionembedding}
        \State Obtain classification matrix \(\mathbf{M} = \mathcal{C}(\mathbf{T})\)
        \State Calculate loss \(\mathcal{L}(\mathbf{M}, \mathbf{I}, \mathbf{L})\) via Eq. \ref{eq:totalloss}
        \State Update parameters: \(\theta_{a}^{*}, \theta_{w}^{*}, \theta_{t}^{*} \leftarrow \arg\min_{\theta} \mathcal{L}\)
    \EndFor
\EndFor
\Statex
\State \textbf{Output:} Laundering predictions \(\mathbf{L}' \leftarrow \arg\max(\mathbf{M}[:, :2])\)
\end{algorithmic} 
\label{alg:crowdmlmodel}
\end{algorithm}

As depicted in Fig. \ref{fig:crowdsourcingmlstructure}, the multi-level transaction embeddings serve as inputs of a unified classifier for multi-task detection and collaborative optimization. Algorithm \ref{alg:crowdmlmodel} details this optimization workflow, taking transaction graph \(\mathcal{G}\), laundering labels \(\mathbf{L}\), and group labels \(\mathbf{I}\) as inputs. Each epoch of iteration begins with partitioning \(\mathcal{G}\) into batches of subgraphs \(\{\mathcal{G}'_i\}\) for training. Within each batch, multi-level transaction embeddings are computed using Eq. \ref{eq:accountembedding}, \ref{eq:weightembedding}, and \ref{eq:transactionembedding}. These embeddings are fed into the unified classifier \(\mathcal{C}\) to yield detection results \(\mathbf{M}\). The multi-task classification loss is then computed and minimized via the Adam optimizer to update model parameters \(\theta_{a}^{*}, \theta_{w}^{*}, \theta_{t}^{*}\).

The loss function employs standard binary cross-entropy for both tasks. The laundering transaction classification loss \(\mathcal{L}_m\) is:
\begin{equation}
    \mathcal{L}_{m} = - \frac{1}{m}\sum_{i=1}^{m} \sum_{j=0}^{1} w_j \cdot \mathbf{L}_i \cdot \log(\mathbf{L}'_i),
\end{equation}
where \(w_j\) is a class weight balancing parameter to mitigate data imbalance, \(\mathbf{L}_i\) denotes the true label for transaction \(e_i\), and \(\mathbf{L}'_i\) is its predicted results. The transaction group prediction loss \(\mathcal{L}_g\) is:
\begin{equation}
    \mathcal{L}_{g} = - \frac{1}{m}\sum_{i=1}^{m} \mathbf{I}_i \cdot \log(\mathbf{I}'_i),
\end{equation}
where \(\mathbf{I}_i\) and \(\mathbf{I}'_i\) represent the true and predicted group labels. The total loss combines both terms:
\begin{equation}\label{eq:totalloss}
    \mathcal{L} = \mathcal{L}_{m} + \lambda \mathcal{L}_{g}.
\end{equation}

Here, \(\lambda\) is a weighting coefficient balancing task importance. This joint loss enables collaborative optimization of shared embeddings. Crucially, the auxiliary task (transaction group detection) leverages known group information to drive intra-group transaction embeddings toward similarity, significantly enhancing generalization for money laundering detection. This is particularly valuable in crowdsourcing laundering scenarios containing diverse transaction patterns. Furthermore, benefiting from GNNs' inherent label propagation mechanism, the end-to-end model exhibits strong semi-supervised capacity, efficiently integrating limited labels with abundant unlabeled data to reduce annotation dependency in cryptocurrency laundering scenarios.

\section{Experiments}
In this section, we evaluate the effectiveness of our method through extensive experiments. Our method detects crowdsourcing laundering by introducing an auxiliary task (transaction group detection), and this is also applicable to general laundering detection scenarios. We evaluate different detection methods under both crowdsourcing and general laundering scenarios separately to answer the following research questions:

\begin{itemize}
    \item RQ1: How effective is our method versus existing methods for crowdsourcing cryptocurrency laundering?
    \item RQ2: How effective is our method versus existing methods for general cryptocurrency laundering (e.g., hack laundering)?
    \item RQ3: Can any component be removed or simplified without significant performance degradation?
\end{itemize}

\subsection{Datasets}

\begin{table}[!t]
\centering
\caption{TRON addresses of suspicious "running points" Telegram chats}
\label{tab:telegram}
\begin{tabular}{c|l}
\toprule
Chat ID & Public TRON Addresses \\ \midrule
1 & TMr93D5eqPbmsNGdZAeZfiABLirL5HZY2B \\ \midrule
\multirow{2}{*}{2} & TJh5zGuMA4jLeAZNbAFFQ4t2EkacYZgxa1 \\ 
    & TPsjfDqBBXoMfHSbUWMqMAW9RktbGrCyCH \\ \midrule
3 & TJh5zGuMA4jLeAZNbAFFQ4t2EkacYZgxa1\\ \midrule
\multirow{2}{*}{4} & TB73gtW1hsTxxA9XUYKLSJKTSHw6Jk53eH\\ 
& TWDchZBmYvTQBeXD4w8rRUowDv5ka8kiFU \\ \midrule
5 & TPsjfDqBBXoMfHSbUWMqMAW9RktbGrCyCH\\ \bottomrule
\end{tabular}
\end{table}

\begin{table}[!t]  
\centering  
\caption{Overview of cryptocurrency laundering datasets}  
\label{tab:datasets_paofen}  
\begin{tabular}{lc|ccc}  
\toprule  
\textbf{Statistic}   & \emph{Tron-USDT} & \emph{Harmony} & \emph{Upbit}  & \emph{IBM-LI}  \\ \midrule  
\# of $\mathcal{A}$               & 1043K     & 340K    & 577K   & 705K  \\   
\# of $\mathcal{T}$        & 2.6M      & 1.8M    & 2.3M   & 7.0M  \\   
\# of $\mathbf{L}$ & 252       & 409     & 58,643 & 1,221 \\ 
\% of $\mathbf{L}$             & 10,529    & 4,573   & 40     & 5,733 \\     
\# of $\mathbf{I}$     & 18,997   & \text{N/A} & \text{N/A} & \text{N/A} \\  
\# of Groups  & 193     &  \text{N/A} & \text{N/A} & \text{N/A} \\ \bottomrule   
\end{tabular}  
\end{table}  

Corresponding to two laundering scenarios, we construct two types of datasets:

\begin{itemize}
  \item \textbf{Crowdsourcing laundering dataset}: Built from suspicious "running points" Telegram chats and transaction analysis. This dataset leverages TRON's staking delegation relationships as transaction group labels to validate full performance of MCCLD algorithm.
  
  \item \textbf{General laundering datasets}: Contain real-world cryptocurrency hacker laundering cases and synthetic AML simulation data. These datasets lack transaction group information and are primarily used to evaluate the algorithm’s generalization capability in general scenarios.
\end{itemize} 


\textbf{Crowdsourcing laundering transaction graph.} Telegram is a primary recruitment channel for crowdsourcing laundering gangs\cite{UNODC2024}. Through keyword searches (e.g., "black USDT", "heiu" and "paofen"), we extract the top 5 most active "running points" chats, all of which remain operational to date. From these chats, we obtain available transaction records and related TRON addresses, which are made public in chats and used for further advertising. Involved TRON addresses are presented in Table \ref{tab:telegram}. Notably, Chats 2 and 3 exhibit same TRON addresses, which should be controlled by the same criminal group. By tracing historical transactions of these addresses and selecting the most frequent transaction interval (blocks 50763000 to 50793000) across above five chats, we construct the transaction graph based on fully TRON transactions. For laundering transaction labeling, we obtain the laundering labels from public risk annotations of blockchain explorer -- Oklink\footnote{https://www.oklink.com/}. For transaction group labeling, we discover the mechanism of resource staking and delegation that is commonly used to reduce transaction fee either crowdsourcing laundering gangs and normal users, and we treat each connected component in delegation relationship network as a transaction group in TRON. As this laundering activity occurred on TRON using primarily USDT, we named the resulting graph \emph{TRON-USDT}, and its statistical overview exhibited in Table \ref{tab:datasets_paofen}.

\textbf{General money laundering transaction graphs.} We construct general laundering graphs using two open-source datasets: \emph{Upbit} \cite{Wu2024} and \emph{Harmony} \cite{LiG2025}. These datasets track real-world cryptocurrency hacker laundering cases (Upbit exchange hack and Harmony cross-chain bridge hack) through on-chain transaction records. In addition, we incorporat a public available multi-currency synthetic money laundering transaction dataset \emph{IBM-LI} \cite{Altman2023}. Notably, these datasets lack explicit transaction group information, so we exclude the transaction group loss metric when evaluating MCCLD on these graphs, so as to validate the algorithm’s generalization capability for general cryptocurrency laundering scenarios. Table \ref{tab:datasets_paofen} summarizes statistic information of all graphs used in this paper.

\subsection{Setup}  
\textbf{Evaluation Metrics.} Considering the inherent class imbalance in laundering transaction classification task, we employ two widely used evaluation metrics for imbalanced classification: F1-score and AUC. The F1-score balances precision and recall, effectively measuring the model’s overall ability to detect both of positive and negative samples. The AUC value reflects the model’s discriminative power across different classification thresholds.  

\textbf{Baseline Algorithms.} The proposed MCCLD framework is compared against two types of cryptocurrency laundering detection methods: (i) Subgraph-based laundering detection algorithms, including GMPA \cite{LiG2025} and AntiBenford \cite{Chen2022}; (ii) Transaction-based laundering detection algorithms, primarily supervised graph neural network models such as Graph Convolutional Network (GCN), Graph Attention Network (GAT), GraphSAGE, and Principal Neighbor Aggregation (PNA).  

\textbf{Parameters Setting.} For MCCLD, the node embedding output dimension is set to 64, the GNN layer count is 2, the learning rate is 0.006, and the loss weight coefficient is 0.5. Baseline GNN algorithms retain their default parameter settings from prior work\footnote{https://github.com/IBM/Multi-GNN}.  


\subsection{Crowdsourcing Laundering Detection}  

\begin{figure}[!t]  
\centering  
\includegraphics[width=0.35\textwidth]{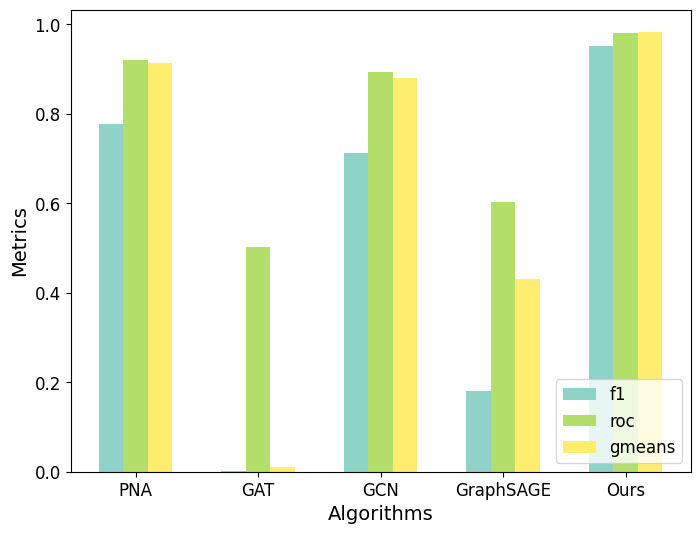}  
\caption{Detection performance of algorithms on the crowdsourcing laundering graph}  
\label{fig:paofen_metrics}  
\end{figure}

Evaluation metrics were computed by comparing model predictions against the ground truth of laundering transactions. Results in Fig. \ref{fig:paofen_metrics} show MCCLD's superior performance: the F1-score of 0.95 indicates precise detection with low false positives/negatives, and the AUC of 0.98 reflects excellent discrimination between laundering and legitimate transactions. Compared to baselines, MCCLD achieves significant improvements—average increases of 53.4\% in F1-score and 25.2\% in AUC—highlighting its leading performance in crowdsourcing laundering detection.  

Notably, while GAT and GraphSAGE excel in traditional finance laundering detection \cite{Chai2023,Cheng2023}, their performance degrades in cryptocurrency. This is attributable to the unique topology of cryptocurrency networks: accounts usually connect to high-centrality nodes (e.g., exchanges, smart contracts), and neighbor diversity far exceeds that in traditional finance. Such complex topology causes "attention drift" in GAT, making the model focus on  relationships with high weights but not related to money laundering. Similarly, GraphSAGE’s neighbor sampling fails to preserve laundering-relevant neighborhood information, impairing its performance. In contrast, GCN maintains stable performance via full-neighborhood aggregation, and PNA enhances adaptability to complex networks via multi-scale aggregation functions, demonstrating superior representation learning for cryptocurrency laundering transactions.

\subsection{General Money Laundering Detection}

Experimental results of various algorithms on general money laundering transaction datasets are presented in Table \ref{tab:paofen_comparison}. Benefiting from multi-level transaction embedding fusion, the MCCLD algorithm demonstrates exceptional generalization capability even without transaction group information. Particularly on the \emph{Harmony} dataset, it achieves an F1-score of 0.806 and an AUC of 0.916, outperforming baselines by 36.9\% in average F1-score. In contrast, due to the unsupervised nature, subgraph-based methods consistently yield F1-scores below 0.1, showing significant performance gap versus MCCLD.

\begin{figure}[!t]
\centering
\includegraphics[width=0.35\textwidth]{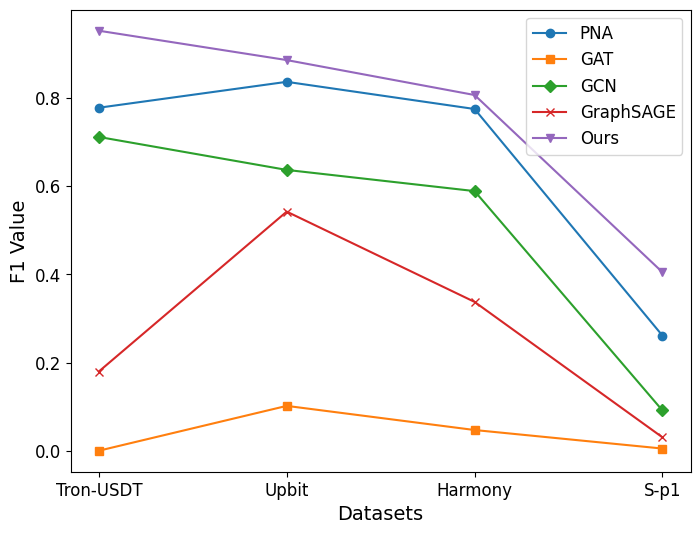}
\caption{Detection performance of algorithms across all datasets}
\label{fig:paofen_datasets}
\end{figure}

Fig. \ref{fig:paofen_datasets} illustrates the F1-score distribution across all datasets, MCCLD exhibits the most pronounced advantage in crowdsourcing laundering scenario, strongly validating its effectiveness for such activities. Notably, all algorithms experience significant performance degradation on the synthetic dataset \emph{IBM-LI}, which is likely attributed to oversimplified laundering patterns in this dataset that lack differentiation from legitimate transactions. Despite this, MCCLD elevates the state-of-the-art F1-score \cite{Altman2023} on \emph{IBM-LI} from 0.273 to a new benchmark.

Algorithms display varying adaptability across different datasets: compared to the crowdsourcing laundering dataset \emph{Tron-USDT}, algorithms of PNA, GraphSAGE, and GAT show improved performance on \emph{Upbit} and \emph{Harmony}, while MCCLD and GCN exhibit the opposite trend. This divergence stems from biased neighbor aggregation designs in PNA, GraphSAGE, and GAT, favoring homogeneous neighborhoods. Conversely, GCN's full-neighborhood aggregation aligns better with diverse neighborhood environments. Furthermore, when group information is absent, MCCLD processes neighborhood information similarly to GCN, influencing its cross-dataset performance variations.

\begin{table*}[!t]
\centering
\caption{Detection results of algorithms on general money laundering datasets} 
\label{tab:paofen_comparison}
\begin{tabular}{ll|cc|cc|cc}
\toprule
\multirow{2}{*}{Method Type} & \multirow{2}{*}{Dataset} & \multicolumn{2}{c}{\emph{Upbit}} & \multicolumn{2}{c}{\emph{Harmony}} & \multicolumn{2}{c}{\emph{IBM\_LI}} \\ \cline{3-8}
& & \textbf{F1} & \textbf{AUC} & \textbf{F1} & \textbf{AUC} & \textbf{F1} & \textbf{AUC} \\ \midrule
\multirow{2}{*}{Subgraph-Based} & GMPA & 0.2927 & 0.8095 & 0.1825 & 0.5657 & 0.0713 & 0.6568 \\ 
& Anti-Benford & 0.2640 & 0.5826 & 0.0199 & 0.4999 & 0.0083 & 0.4967 \\ \midrule
\multirow{5}{*}{Transaction-Based} & PNA & 0.836 & 0.892 & 0.774 & 0.911 & 0.261 & 0.700 \\
& GAT & 0.102 & 0.502 & 0.047 & 0.572 & 0.005 & 0.501 \\ 
& GCN & 0.636 & 0.759 & 0.588 & 0.908 & 0.091 & 0.528 \\ 
& GraphSAGE & 0.542 & 0.669 & 0.337 & 0.852 & 0.030 & 0.508 \\ \cline{2-8}
& \textbf{MCCLD} & \textbf{0.885} & \textbf{0.923} & \textbf{0.806} & \textbf{0.916} & \textbf{0.405} & \textbf{0.737} \\ \bottomrule
\end{tabular}
\end{table*}

\subsection{Ablation Study}

Integrating transaction group information to assist laundering detection constitutes the core innovation of MCCLD. This section investigates the impact of transaction group information on detection performance and also explores the algorithm's label dependency under scarce supervision scenarios.

\textbf{Impact of group information.} Among all datasets, only \emph{Tron-USDT} contains native transaction group information (staking relationships). To derive group information for other datasets, we employ the Louvain algorithm \cite{Blondel2008} and GMPA algorithm \cite{LiG2025} to partition transaction graphs, and use the resulting subgraphs as transaction groups for auxiliary supervision. To prevent all-one group vectors (invalid when fully partitioned), we filtered subgraphs ranging from 2-10000 nodes.

\begin{table*}[!t]
\centering
\caption{The performance of MCCLD under different auxiliary task settings} 
\label{tab:abl_group}
\begin{tabular}{lcc|cc|cc|cc}
\toprule
\multirow{2}{*}{Group Info Source} & \multicolumn{2}{c}{\emph{Tron-USDT}} & \multicolumn{2}{c}{\emph{IBM-LI}} & \multicolumn{2}{c}{\emph{Upbit}} & \multicolumn{2}{c}{\emph{Harmony}} \\ \cline{2-9}
& \textbf{F1} & \textbf{AUC} & \textbf{F1} & \textbf{AUC} & \textbf{F1} & \textbf{AUC} & \textbf{F1} & \textbf{AUC} \\ \midrule
No Group Info & 0.8533 & 0.971 & 0.405 & 0.737 & \textbf{0.885} & 0.923 & \textbf{0.806} & 0.916 \\
GMPA-Generated Groups & \textbf{0.9562} & \textbf{0.9857} & \textbf{0.6657} & \textbf{0.8849} & 0.7949 & 0.9039 & 0.5915 & 0.8984 \\
Louvain-Generated Groups & 0.9406 & 0.9802 & 0.3851 & 0.7367 & 0.7378 & 0.912 & 0.541 & 0.9746 \\ \bottomrule            
\end{tabular}
\end{table*}

Experiments were conducted on multi-gang datasets (\emph{Tron-USDT}, \emph{IBM-LI}) and single-gang datasets (\emph{Upbit}, \emph{Harmony}). Results (Table \ref{tab:abl_group}) reveal: In multi-gang scenarios, introducing group information via GMPA subgraphs or Louvain communities significantly enhances detection performance, improving F1-scores by 10.29\% and 26.07\% respectively over the no-group baseline. Conversely, in single-gang scenarios, externally derived group information degrades performance. This suggests that graph partitioning provides additional structural information for multi-gang laundering detection but may disrupt the already limited structural single laundering gangs, weakening auxiliary detection capability. Furthermore, GMPA, which partition graphs using transaction semantics, demonstrates superior performance as an auxiliary method for MCCLD. When utilizing GMPA-generated groups, MCCLD consistently outperforms Louvain-based configurations across F1 and AUC metrics. Remarkably, on \emph{Tron-USDT}, MCCLD with GMPA groups achieves performance comparable to the the native staking relationship baseline, validating that high-quality group feature extraction is crucial for the accuracy of money laundering detection.

\begin{figure}[!t]
\centering
\includegraphics[width=0.4\textwidth]{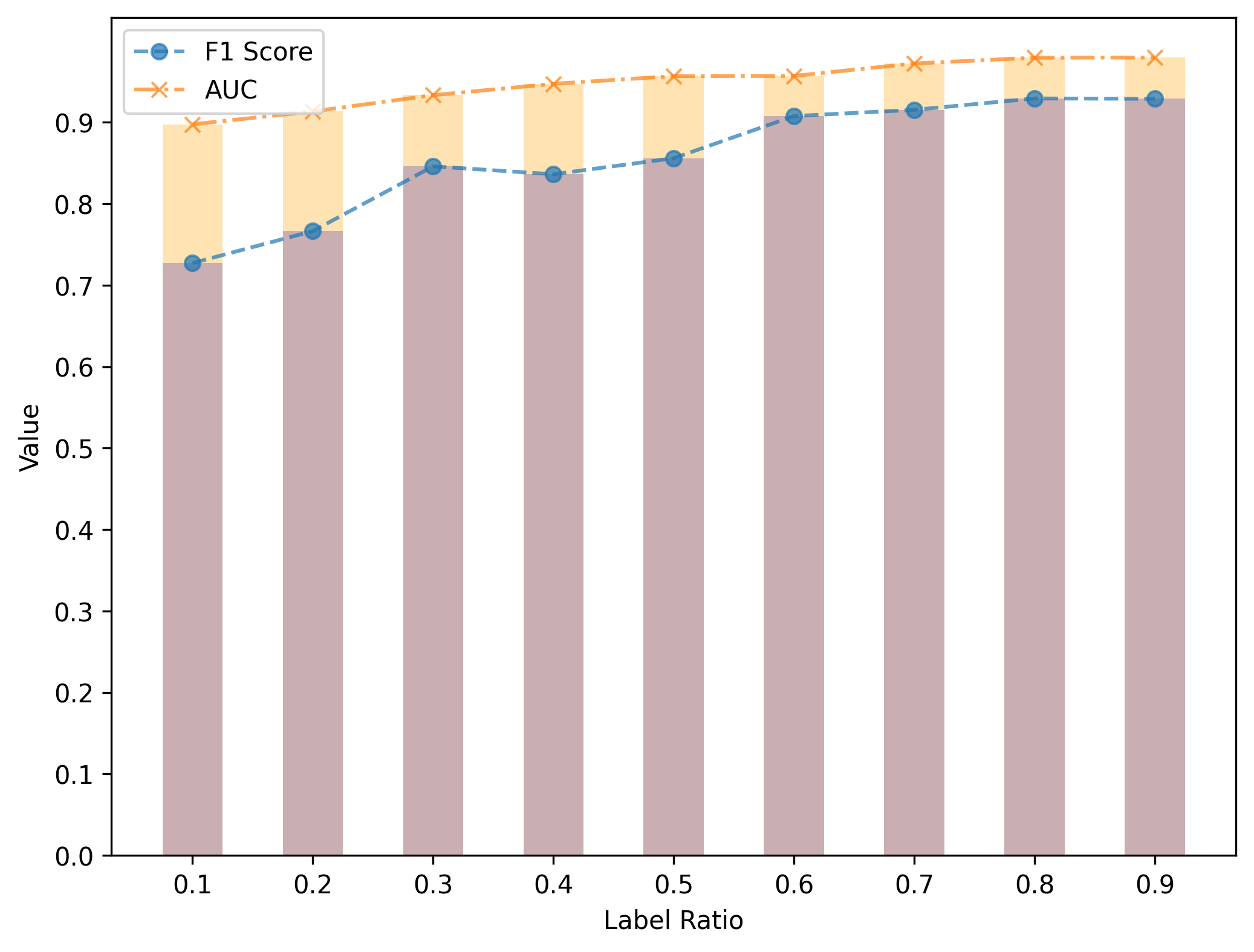}
\caption{The performance of MCCLD on crowdsourcing laundering under varying label ratios}
\label{fig:alaphs_labelratio}
\end{figure}

\textbf{Semi-supervised capability.} Benefiting from GNNs' message passing \cite{Thomas2017}, MCCLD exhibits excellent semi-supervised capability. As shown in Fig. \ref{fig:alaphs_labelratio}, F1-score of MCCLD remains above 70\% across all label ratios (0.1-0.9) on \emph{Tron-USDT}. Under extreme low-label conditions (only 10\% training labels), it still achieves an F1-score of 72\%, confirming effectiveness in the scenarios of scarce annotations.

\section{Conclusion}
This paper studies a new form of crowdsourcing cryptocurrency laundering, which exhibits distinctive characteristics compared to traditional money laundering: diversified transaction patterns and polycentric structure. To address these challenges, we proposes the Multi-Task Collaborative Cryptocurrency Laundering Detection (MCCLD) framework. This framework constructs an end-to-end graph neural network that integrates account and transaction attributes to generate multi-level transaction embeddings. Its core innovation lies in the cooperative design of primary and auxiliary tasks: laundering transaction detection (primary) and transaction group detection (auxiliary) share feature embeddings and classifiers. Through auxiliary task constraints that force intra-group transaction embeddings to converge, the framework significantly enhances generalization capability for diverse patterns within the same laundering group. Experimental results demonstrate MCCLD's effectiveness in real-world crowdsourcing laundering scenarios while maintaining robust generalization in conventional laundering scenarios (e.g., hacker laundering) without explicit group information.

\balance
\bibliographystyle{ACM-Reference-Format}
\bibliography{sample-base}










\end{document}